\documentstyle[prl,aps,preprint]{revtex}

\begin{document}
\tighten

\title{{DEEPLY-VIRTUAL COMPTON SCATTERING}
\thanks {This work is supported in part by funds provided by the
U.S.  Department of Energy (D.O.E.) under cooperative agreements
\#DF-FC02-94ER40818 and \#DOE-FG02-93ER-40762.}}

\author{Xiangdong Ji}

\address{Department of Physics \\
University of Maryland \\
College Park, Maryland 20742 \\
and \\
Center for Theoretical Physics \\
Laboratory for Nuclear Science \\
and Department of Physics \\
Massachusetts Institute of Technology \\
Cambridge, Massachusetts 02139 \\
and \\
Institute for Nuclear Theory \\
University of Washington \\
Seattle, Washington 98195 \\
{~}}

\date{U. of MD PP\#97-026 ~~~ MIT-CTP-2568 ~~~ hep-ph/9609381 ~~~ September 1996}

\maketitle

\begin{abstract}
We study in QCD the physics of deeply-virtual Compton 
scattering (DVCS)---the virtual Compton process in 
the large $s$ and small $t$ kinematic region. 
We show that DVCS can probe a new type of
{\it off-forward} parton distributions. We derive 
an Altarelli-Parisi type of evolution equations 
for these distributions. We also derive their sum rules
in terms of nucleon form-factors of the twist-two
quark and gluon operators. In particular, we find
that the second sum rule is related to fractions
of the nucleon spin carried separately by quarks and 
gluons. We estimate the cross section for 
DVCS and compare it with the accompanying 
Bethe-Heitler process at CEBAF and HERMES kinematics.
  
\end{abstract}
\pacs{xxxxxx}

\narrowtext

\section{Introduction}
The Compton process, which refers to elastic scattering of
a photon off a charged object, has played an 
important role
in the history of Quantum Electrodynamics: It provided one
of the early evidences that the electromagnetic wave is quantized, 
and hence has the nature of particles\cite{compton}. The role  
of the Compton process in studying the structure of hadrons
has been explored since 50's, when Low's low-energy theorems \cite{low},
analogous to the well-known Thomson cross section,
are derived. Those theorems assert, for instance, that
at sufficiently low energy the spin-dependent part of 
the Compton amplitude is 
determined by the anomalous magnetic moment of a composite system.
Going to higher-order terms in the low-energy expansion,
one finds the electric and magnetic polarizabilities\cite{pol1}. 
In recent years, experimental and theoretical works 
in measuring and 
understanding the polarizabilities of the nucleon and pion
have flourished\cite{pol2}.

Generally speaking, however, the Compton process on a composite
system is quite complicated. When a point-like
constituent absorbs an incoming photon, the system becomes
excited. As it propagates in time, the system eventually emits 
a photon and comes back to the ground state. Quantum-mechanical 
propagation of a composite system is difficult to 
handle theoretically, except
in spacial kinematic regions. The low-energy theorems 
exist for the Compton scattering on the nucleon
because the intermediate propagation 
at low energy is dominated by the nucleon itself\cite{low}. 
Another kinematic region known to have simple scattering mechanism
is where the $t$-channel momentum transfer is large, i.e., 
the nucleon has a large recoil\cite{farrar}. In this
case, perturbative Quantum Chromodynamics can be
used to understand the intermediate propagation. In fact, 
according to the so-called the power counting rule\cite{brodsky0}, 
the most important intermediate states are those created 
from three valence quarks through 
hard-gluon exchanges. Simple as it may be,
one still has to compute hundreds of Feynman
diagrams to obtain the scattering cross section. 

The purpose of this paper is to study the Compton scattering 
by a virtual photon in a special kinematic limit.
Assuming the virtual photon is generated by inelastic 
lepton scattering, we are interest in the Bjorken limit, i.e., 
the energy and momentum of the virtual 
photon going to infinity at the same rate. We shall 
call the process deeply-virtual Compton scattering (DVCS). 
As we shall discuss in the next section, the basic 
mechanism for DVCS is a quark absorbing the virtual photon,
immediately radiating a real photon and falling back to the 
nucleon ground state. Thus the physics of DVCS is quite 
simple. 

Our interest in studying DVCS is generated from the fact that
it offers a way to measure the off-forward parton
distributions (OFPDs), a new type of parton distributions that 
generalize the usual parton distributions 
and the nucleon form factors. When taking moments
of OFPDs, one gets form factors of the spin-$n$, twist-two
quark and gluon operators. When going to the forward limit, 
OFPDs become the usual quark and gluon distributions. 
Because the spin-two, twist-two operators are part of 
the energy-momentum tensor of QCD and because the form
factors of the energy-momentum tensor contain information
about the quark and gluon contributions to the 
nucleon spin, DVCS may provide a novel way to measure the fraction 
of the nucleon spin carried by the 
quark orbital angular momentum, a subject of great current
interest\cite{e143}.

The presentation of the paper goes like this. In Section II, 
we consider DVCS in QCD at the tree level. We identify the off-forward
parton distributions from the Compton amplitude and then study some
simple aspects of the distributions, such as sum rules.
In Section III, we study the leading-log evolution of 
the OFPDs. The results are presented in the form of 
generalized Altarelli-Parisi
equations. In sections IV, we work out the cross sections
for DVCS and the accompanying Bethe-Heitler process 
and their interference. To be complete, we consider 
all cases including unpolarized, double-spin and single-spin 
processes. Some estimates are given at the CEBAF and 
HERMES kinematics.
The final section contains comments and discussions.

A preliminary account of the DVCS process is discussed in 
a Letter paper by this author \cite{ji}. Subsequently,
Radyushkin studied the scaling limit at $\Delta^2=0$
from a different angle \cite{r1}. Actually, processes 
similar to DVCS were first considered by Geyer et. al.
in studying the anomalous dimensions of light-ray
operators \cite{geyer}, where the ``interpolating
functions'' were introduced. The evolution of these 
functions is found to 
interpolate the Brodsky-Lepage and Altarelli-Parisi equations.   
Similar objects were also considered by Jain and Ralston
in the context of studying the violation of 
helicity selection rule and the effects of transverse momentum
\cite{ral}. 

\section{Deeply-virtual compton scattering at leading order}

This section is devoted to studying the virtual Compton 
scattering in deep-inelastic kinematics and
at the leading order in perturbative QCD. The main result
is that DVCS is dominated by single-quark scattering,
and therefore the amplitude can be expressed in terms
of the off-forward parton distributions. We also study sum rules 
of these distributions and show that the second sum rule 
is related to the total quark (gluon) contribution to the 
spin of the nucleon.  

We picture the virtual Compton 
scattering in Fig. 1, where
a nucleon of momentum $P^\mu$
absorbs a virtual photon of momentum $q^\mu$, producing 
an outgoing real photon of 
momentum $q'^\mu=q^\mu - \Delta^\mu $ and a recoil 
nucleon of momentum $P'^\mu=P^\mu+\Delta^\mu$.
We focus on  
the deeply-virtual kinematic region of $q^\mu$, namely, 
the Bjorken limit: $Q^2=-q^2\rightarrow \infty$,  
$P\cdot q\rightarrow\infty$, and $Q^2/P\cdot q$ finite.
In this region, the quark that absorbs the virtual 
photon becomes highly virtual and hence propagates 
perturbatively. The simplest mechanism to 
form the Compton final state is for the quark to promptly 
radiate a real photon and fall back to the nucleon ground
state. This ``hand-bag" subprocess is shown in Fig. 2(a).
 
In QCD, more complicated ``tree" subprocesses
are possible. By tree, we mean perturbative diagrams 
in which every vertex is next to the nucleon blob. 
For instance, the highly-virtual quark can interact with 
gluon fields in the nucleon, as shown in Fig. 2(b); 
or it can transfer its virtuality to another 
quark through one-gluon exchanges, as shown in Fig. 2(c). 
A detailed calculation shows that both subprocesses
are suppressed by $1/Q^2$ relative to the hand-bag
diagram, except
when the polarization of the gluon in Fig. 2(b) is
longitudinal. Fortunately, in the light-cone gauge $A^+=0$, 
the longitudinally-polarized gluons do not contribute by definition.
Thus in the deeply-virtual limit, the single-quark
process indeed dominates the Compton scattering.

Of course, one can decorate the hand-bag 
with radiative loops, such as
those shown in Figs. 2(d) and 2(e). Those diagrams in certain
kinematic region produce leading-log corrections to 
the simple hand-bag, as we shall discuss in the next
section. In other kinematic regions, they
give rise to order-$\alpha_s(Q^2)$ radiative corrections
and hence can be ignored in the deeply-virtual limit.
One exception is the gluon vertex correction
for the real photon in Fig. 2(e), where as the two quark lines
have momenta predominantly parallel to the photon momentum, 
the diagram has an infrared divergence reflecting the non-perturbative
photon wave function. 
Fortunately, a simple calculation shows that the contribution
in this kinematic region is suppressed by $1/Q^2$ relative to
the hand-bag due to the hard-gluon propagator.
 
Thus, in the remainder of this section we concentrate
solely on the dominant subprocess in Fig. 2(a). 

According to Feynman rules, the hand-bag diagram corresponds to
the Compton amplitude, 
\begin{eqnarray} 
T^{\mu\nu} &=& i \int {d^4k\over (2\pi)^4}
       ~{\rm Tr}~\{[\gamma^\nu {i\over 
       \not\! k-\alpha \not\! \Delta +\not\! q +i\epsilon}
        \gamma^\mu \nonumber \\ &&
        + \gamma^\mu {i\over \not\! k
       + (1-\alpha)\not\! \Delta - \not\! q +i\epsilon}
          \gamma^\nu]M(k)\} \ , 
\end{eqnarray}
where $\mu$ and $\nu$ 
are the polarization indices of the virtual and real photons and 
$M(k)$ is a quark density matrix, 
\begin{equation}
       M(k) = \int e^{ikz}d^4 z \langle P'|
   \bar \psi(-\alpha z)\psi((1-\alpha)z)|P\rangle \ , 
\end{equation}
where $0<\alpha<1$ reflects the arbitrariness of the looping
momentum $k^\mu$. To 
proceed further, it is convenient to  
define a special system of coordinates. 
We choose $q^\mu$ and $\bar P^\mu = 
(P+P')^\mu/2$ to be collinear and in the $z$ direction. 
Introduce two light-like vectors, $p^\mu=\Lambda(1,0,0,1)$
and $n^\mu=(1,0,0,-1)/(2\Lambda)$, with $p^2=n^2=0$, $p\cdot n=1$, 
and $\Lambda$ arbitrary. 
We expand other vectors in terms of $p^\mu$, $n^\mu$ and transverse 
vectors,
\begin{eqnarray}
           \bar P^\mu &=& p^\mu + (\bar M^2/2) n^\mu  \ , \nonumber \\
           q^\mu &=& -\xi p^\mu + (Q^2/2\xi) n^\mu  \ ,  \nonumber \\
          \Delta^\mu &=& -\xi(p^\mu-(\bar M^2/2) n^\mu) + \Delta_\perp^\mu
\ ,    \nonumber \\
         k^\mu &=& (k\cdot n)p^\mu + (k\cdot p )n^\mu + k^\mu_\perp \ ,  
\end{eqnarray} 
where $ \bar M^2 = M^2 -\Delta^2/4$ and 
$\xi = (-\bar P\cdot q +\sqrt{(\bar P\cdot q)^2+Q^2\bar M^2})/\bar M^2$.
The $\xi$ variable is analogous to the Bjorken variable $x_B$
in deep-inelastic scattering and is bounded by 0 and 
$\sqrt{-\Delta^2}/\sqrt{M^2-\Delta^2/4}$.  
We neglect components of four vectors which do not produce large
scalars in the Bjorken limit. Introducing the factor
$\int dx {d\lambda \over 2\pi} e^{i\lambda (x-k\cdot n)}=1$
and integrating over $k^\mu$ and $z^\mu$, we simplify 
the Compton amplitude to, 
\begin{eqnarray}
         T^{\mu\nu}(P,q,\Delta) 
           &&  =  {1\over 2} (g^{\mu\nu} - p^\mu n^\nu - p^\nu n^\mu)
             \int^1_{-1} dx \left({1\over x-\xi/2 + i\epsilon} 
            + {1\over x+\xi/2 - i\epsilon}\right) \nonumber 
              \\ && \times \left[ H(x,\xi,\Delta^2)
                     \bar U(P') {\not\! n} U(P)
             + E(x, \xi,\Delta^2) 
                   \bar U(P') {i\sigma^{\alpha\beta}n_\alpha
              \Delta_\beta \over 2M}U(P) \right] \nonumber \\ &&
              + {i\over 2}\epsilon^{\mu\nu\alpha\beta}
                p_\alpha n_\beta
                 \int^1_{-1} dx \left({1\over x-\xi/2 + i\epsilon} 
            - {1\over x+\xi/2 - i\epsilon}\right)  \nonumber \\ 
            && \times
           \left[ \tilde H(x,\xi,\Delta^2)\bar U(P')
           \not\! n \gamma_5 U(P) + \tilde E(x, \xi, \Delta^2)
                  {\Delta\cdot n\over 2M}\bar U(P')\gamma_5 U(P) \right] \ , 
\end{eqnarray}
where we have chosen $\alpha=1/2$ for symmetrical reason. $H$, $\tilde H$, $E$ and $\tilde E$ 
are off-forward, 
twist-two parton distributions defined through the following light-cone
correlation functions, 
\begin{eqnarray}
 \int  {d\lambda \over 2\pi} e^{i\lambda x}
      \langle P'|\bar\psi(-{\lambda n/ 2})\gamma^\mu
            \psi(\lambda n/2)|P \rangle
        &=& H(x,\xi,\Delta^2) \bar U(P')\gamma^\mu U(P) \nonumber \\
         && + E(x,\xi \Delta^2) \bar U(P'){i\sigma^{\mu\nu}
             \Delta_{\nu}
          \over 2M}U(P) + ...  \ , \nonumber \\
 \int  {d\lambda \over 2\pi} e^{i\lambda x}
      \langle P'|\bar\psi(-{\lambda n/ 2})\gamma^\mu\gamma_5
            \psi(\lambda n/2)|P \rangle
      & =& \tilde H(x,\xi,\Delta^2) 
        \bar U(P')\gamma^\mu \gamma_5 U(P) \nonumber \\
       && + \tilde E(x,\xi,\Delta^2) \bar U(P')
          {\gamma_5\Delta^\mu
         \over 2M}U(P)
        + ...\ ,
\end{eqnarray}
where the ellipses denote higher-twist distributions. 
According to our definition, the initial nucleon and the
active quark have the
longitudinal momentum $1+\xi/2$ and
$x+\xi/2$, respectively. [In 
covariant gauge, the longitudinal gluons produce a gauge 
link ${\rm exp}(-ig\int^{-\lambda/2}_{\lambda/2}n\cdot A(\alpha
n)d\alpha)$ between the two quark fields, 
restoring explicit gauge invariance 
of the light-cone correlations. 
Here we are working in the light-cone gauge $n\cdot A=0$, 
hence the longitudinal gluons and the gauge link never appear. Nonetheless, 
the result shall be taken as implicitly gauge-invariant.]

A systematic counting indicates
that the virtual Compton scattering depends on twelve 
helicity amplitudes \cite{guichon}. The above result implies that only 
four of them survive the Bjorken limit. An quick inspection shows 
all amplitudes with longitudinally-polarized virtual photons 
are sub-leading. For the transversely-polarized virtual photon scattering, 
the emitted photon retains the helicity of the incident photon
at the leading order. 
This selection rule can serve as a useful check that deeply-virtual 
Compton scattering is indeed dominated by the single quark
process. Of the four off-forward parton distributions, 
$H$ and $\tilde H$ conserve the nucleon helicity, while
$E$ and $\tilde E$
flip the nucleon helicity.

The off-forward parton distributions just defined have
characters of both the ordinary parton distributions and
nucleon form factors. In fact, in the limit of
$\Delta^\mu \rightarrow 0$, we have 
\begin{equation}
     H(x,0,0) = q(x),~~~ \tilde H(x,0,0) = \Delta q(x),
\end{equation}
where $q(x)$ and $\Delta q(x)$ are quark and quark helicity
distributions, defined through similar light-cone
correlations\cite{jaffeji}. On the other hand, forming 
the first moment of the new distributions, one gets
the following sum rules \cite{ji,ral}, 
\begin{eqnarray}
     \int^1_{-1} dx H(x,\xi,\Delta^2) &=& F_1(\Delta^2) \ ,  \nonumber \\
     \int^1_{-1} dx E(x,\xi,\Delta^2) &=& F_2(\Delta^2) \ ,  \nonumber \\
     \int^1_{-1} dx \tilde H(x,\xi,\Delta^2) &=& G_A(\Delta^2) \ , 
           \nonumber \\ 
     \int^1_{-1} dx \tilde E(x,\xi,\Delta^2) &=& G_P(\Delta^2) \ . 
\end{eqnarray}
where $F_1(\Delta^2)$ and $F_2(\Delta^2)$ are the Dirac and Pauli form factors
and $G_A(\Delta^2)$ and $G_P(\Delta^2)$ are the axial-vector and 
pseudo-scalar form factors. [Usually the argument of form factors 
is the positive $-\Delta^2$. Here we omit the minus sign for simplicity.]

The second moment of the parton distributions
is relevant to the spin
structure of the nucleon. To see this, we first
write down the angular momentum operator in QCD as the sum of 
quark and gluon contributions \cite{ji},
\begin{equation}
    \vec{J}_{\rm QCD} = \vec{J}_{q} + \vec{J}_g \ , 
\end{equation}
where
\begin{eqnarray}
     \vec{J}_q &=& \int d^3x ~\vec{x} \times \vec{T}_q \nonumber \\
                 &=& \int d^3x ~\left[ \psi^\dagger 
     {\vec{\Sigma}\over 2}\psi + \psi^\dagger \vec{x}\times (-i\vec{D})\psi\right]
     \ ,  \nonumber \\
     \vec{J}_g &=& \int d^3x ~\vec{x} \times (\vec{E} \times \vec{B}) \ . 
\end{eqnarray}
Here $\vec {T}_q$ and $\vec{E}\times \vec{B}$ are 
the quark and gluon momentum densities, respectively. $\vec{\Sigma}$ is
the Dirac spin-matrix and $\vec{D}= \vec{\partial}+
ig\vec{A}$ is the covariant derivative.
By an analogy with the magnetic moment, one can get 
the separate quark and gluon contributions to the nucleon spin 
if the form factors of the momentum density, or equivalently
the energy-momentum tensor of QCD, are known at zero momentum transfer. 
Using Lorentz covariance
and other symmetry principles, one can write down four
form-factors separately for quark and gluon parts 
of the energy-momentum tensor,
\begin{eqnarray}
      \langle P'| T_{q,g}^{\mu\nu} |P\rangle
       &=& \bar U(P') \Big[A_{q,g}(\Delta^2)
       \gamma^{(\mu} \bar P^{\nu)} +
   B_{q,g}(\Delta^2) \bar P^{(\mu} i\sigma^{\nu)\alpha}\Delta_\alpha/2M 
    \nonumber \\
   &&  +  C_{q,g}(\Delta^2)(\Delta^\mu \Delta^\nu - g^{\mu\nu}\Delta^2)/M
   + \bar C_{q,g}(\Delta^2) g^{\mu\nu}M\Big] U(P) \ , 
\end{eqnarray}     
where again $\bar P^\mu=(P^\mu+{P^\mu}')/2$, $\Delta^\mu = {P^\mu}'-P^\mu$, 
and $U(P)$ is the nucleon spinor. Substituting the above
into the nucleon matrix element of $\vec{J}_{q,g}$, one 
finds fractions of the nucleon spin carried by quarks, $J_q$,
and gluons, $J_g$, 
\begin{eqnarray}
   J_{q,g} &=& {1\over 2}[A_{q,g}(0)+B_{q,g}(0)]\ ; \nonumber \\ 
  J_q + J_g &=& { 1\over 2}\ . 
\end{eqnarray}

According to the definition, the second moment 
of off-forward parton distributions yields the form factors 
of the energy-momentum tensor,
\begin{equation}
     \int^1_{-1} dx x [H(x, \xi, \Delta^2) +
       E(x, \xi, \Delta^2) ]
     = A(\Delta^2) + B(\Delta^2) \ , 
\end{equation} 
where luckily the $\xi$ dependence, or $C_q(\Delta^2)$
contamination, drops out. Extrapolating the sum rule 
to $\Delta^2=0$, one gets $J_{q,g}$. Note that only in this 
special application, we are interested in $\Delta^2\rightarrow 0$ 
limit. In general discussions of DVCS, such limit
is of course not necessary.

By forming still higher moments, one obtains form 
factors of the twist-two 
operators of spin greater than 2. In general, there
are many form-factors for each of the tensor operators; however,
only special combinations of them appear in moments
of the OFPDs. The relative weighting of the different form factors
is determined by the variable $\xi$.

\section{Leading-Log Evolution of Off-Forward Parton Distributions}

In this section, we study the leading-log evolution of
the off-forward parton distributions. Like the 
leading-log evolution of the usual parton distributions
(Altarelli-Parisi equation \cite{ap}),
there are many approaches to calculate it. Here 
we use the momentum-space Feynman diagram technique. For simplicity, 
our calculation is done in the light-cone gauge, although 
one is free to work entirely in covariant gauge. While
the result of Subsection A has appeared in different 
forms in the literature before \cite{geyer,braun,r2}, 
the result of Subsection B is new.  

Evolution of the helicity-independent and helicity-dependent 
distributions are different, hence we treat the two 
cases separately. 

\subsection{Evolution of Parton-Helicity-Independent Distributions}

In this subsection, we consider evolution of  
helicity-independent off-forward parton distributions. 
We use a generic notation $E_{S,NS}(x,\xi,\Delta^2,Q^2)$ to denote 
singlet and non-singlet quark density, 
\begin{equation}
      E_{S,NS}(x,\xi,\Delta^2,Q^2) ={1\over 2} \int {d\lambda \over 2\pi}
          e^{i\lambda x}\langle P'|\bar \psi(-{\lambda \over 2} n)
  \not\! n \psi({\lambda \over 2} n) |P\rangle \ , 
\end{equation}
where the flavor indices and the gauge link have been ignored, 
and $E_G(x,\xi,\Delta^2,Q^2)$ to denote gluon distribution, 
\begin{equation}
      E_G(x, \xi,\Delta^2, Q^2)
    = -{1 \over 2x} \int { d\lambda\over 2\pi}e^{i\lambda x}
  \langle P'|F^{\mu\alpha}(-{\lambda\over2}n)
        F^{\nu}_{~\alpha}({\lambda\over2}n)|P\rangle n_\mu n_\nu \ . 
\end{equation}
Since gluons are bosons, we have $E_G(-x) = -E_G(x)$.
The support for the light-cone variable $x$ can be studied as in 
Ref. \cite{jaffe0} and 
is $-1<x<1$.  
Since the evolution equations are independent of $\Delta^2$, we 
will omit the variable in the following equations.

The evolution of the non-singlet quark density at $x>\xi/2$, 
where both quark lines represent quarks, takes the form, 
\begin{equation}
        { D_QE_{NS}(x,\xi,Q^2)\over D\ln Q^2} = {\alpha_s(Q^2) \over 2\pi}
           \int^1_x  {dy\over y} P_{NS}({x\over y}, 
       {\xi\over y}) E_{NS}(y,\xi,Q^2) \ , 
\label{ap}
\end{equation}
where 
\begin{equation}
     {D_Q\over D\ln Q^2} = {d\over d\ln Q^2}
       - {\alpha_s(Q^2)\over 2\pi}C_F
      \left[{3\over 2} + \int^x_{\xi/2} {dy\over
        y-x-i\epsilon} + \int^x_{-\xi/2}{dy \over
        y-x-i\epsilon}\right] \ . 
\end{equation}
The parton splitting function is calculated according to 
Fig. 3(a),
\begin{equation}
            P_{NS} (x,\xi) =C_F{x^2 + 1 - \xi^2/2 \over
(1-x+i\epsilon)(1-\xi^2/4)}\ , 
\end{equation}
where $C_F=4/3$ for SU(3) color group.
The end-point singularity is cancelled by the divergent integrals
in $D_Q/D\ln Q^2$. 
 Obviously, when $\xi=0$, 
the splitting function becomes the usual Altarelli-Parisi
evolution kernel. 
For $-\xi/2 < x < \xi/2$ where one of the two quark lines
represents a quark and the other represents an antiquark, 
the evolution takes the form, 
\begin{equation}
        { D_QE_{NS}(x,\xi,Q^2)\over D\ln Q^2} = {\alpha_s(Q^2) \over 2\pi}
           \left[\int^1_x  {dy\over y} P'_{NS}({x\over y}, {\xi\over y})
            - \int^x_{-1}{dy\over y} P'_{NS}({x\over y}, -{\xi\over y})\right] 
        E_{NS}(y,\xi,Q^2) \ , 
\end{equation}
where 
\begin{equation}
        P'_{NS}(x, \xi) = C_F{x+\xi/2\over 
          \xi(1+\xi/2)}(1+{\xi\over 1-x+i\epsilon}) \ . 
\end{equation}
When $\xi=2$, shifting the variable $x\rightarrow x-1$
and then scaling the $x$ by a factor of 2, one finds that 
the evolution equation becomes the Brodsky-Lepage 
evolution equation for 
the pion wavefunction\cite{bl}.  For $x<-\xi/2$ where both quark
lines represent antiquarks, the evolution takes 
the same form
as Eq. (\ref{ap}), apart from the replacement 
$\int^1_x \rightarrow -\int^x_{-1}$. 

The evolution of the singlet-quark density mixes with that of the
gluon density. For $x>\xi/2$, the coupled evolution takes the form,
\begin{eqnarray}
        { D_QE_{S}(x,\xi,Q^2)\over D\ln Q^2} &= &{\alpha_s(Q^2) \over 2\pi}
          \int^1_x  {dy\over y} \left[ P_{SS}({x\over y}, 
 {\xi\over y}) E_S(y,\xi,Q^2)
      + 2n_F  P_{SG}({x\over y}, {\xi\over y}) E_G(y,\xi,Q^2)\right] \ , 
    \nonumber \\
        { D_GE_{G}(x,\xi,Q^2)\over D\ln Q^2} &= &{\alpha_s(Q^2) \over 2\pi}
          \int^1_x  {dy\over y} \left[ P_{GS} ({x\over y}, {\xi\over y}) 
     \times   {1\over 2}\left( E_S(y,\xi,Q^2) - E_S(-y,\xi,Q^2)\right) \right.
    \nonumber \\ 
     && ~~~~~~~~~~~~~\left.+ ~ P_{GG}({x\over y}, {\xi\over y})
    E_G(y,\xi,Q^2)\right]
      \ , 
\label{ap1}
\end{eqnarray}
where $n_F$ is the number of quark flavors and
\begin{equation}
     {D_G\over D\ln Q^2} = {d\over d\ln Q^2}
       - {\alpha_s(Q^2)\over 2\pi}C_A
      \left[{11\over 6} - {n_F\over 3C_A}
        + \int^x_{\xi/2} {dy\over
        y-x-i\epsilon} + \int^x_{-\xi/2}{dy \over
        y-x-i\epsilon}\right] \ . 
\end{equation}
The evolution kernels are,
\begin{eqnarray}
      P_{SS}(x, \xi) &=& P_{NS}(x, \xi) \ , \nonumber \\
      P_{SG}(x, \xi) &=& T_F {x^2 + (1-x)^2 -\xi^2/4 \over 
       (1-\xi^2/4)^2}  \ , \nonumber \\
      P_{GS}(x, \xi) &=& C_F {1+(1-x)^2 - \xi^2/4 
       \over x(1-\xi^2/4)} \ ,    \nonumber \\
      P_{GG}(x, \xi) &=& C_A {(x^2-\xi^2/4) \over x(1-\xi^2/4)^2}
               \left[1+{2(1-x)(1+x^2) \over x^2-\xi^2/4} 
        + {1+x-\xi^2/2\over 1-x + i\epsilon}\right] \ , 
\end{eqnarray}
where $T_F=1/2$ and $C_A = 3$. As $\xi\rightarrow 0$, we again
get the usual Altarelli-Parisi evolution kernels. 

For $-\xi/2<x<\xi/2$, we have 
\begin{eqnarray}
        { D_Q E_{S}(x,\xi,Q^2)\over D\ln Q^2} &=& {\alpha_s(Q^2) \over 2\pi}
           \left[\int^1_x  {dy\over y} P'_{SS}({x\over y}, {\xi\over y})
            - \int^x_{-1}{dy\over y} P'_{SS}({x\over y}, -{\xi\over y})\right] 
        E_{S}(y,\xi,Q^2) \nonumber \\
       && + {\alpha_s(Q^2) \over 2\pi} 2n_f
           \left[\int^1_x  {dy\over y} P'_{SG}({x\over y}, {\xi\over y})
            - \int^x_{-1}{dy\over y} P'_{SG}({x\over y}, -{\xi\over y})\right] 
         E_{G}(y,\xi,Q^2) \ , \nonumber \\
        { D_G E_{G}(x,\xi,Q^2)\over D\ln Q^2} &=& {\alpha_s(Q^2) \over 2\pi}
           \left[\int^1_x  {dy\over y} P'_{GS}({x\over y}, {\xi\over y})
            - \int^x_{-1}{dy\over y} P'_{GS}({x\over y}, -{\xi\over
       y})\right] \nonumber \\ &&
      ~~~~~~~~~~~~~~~ \times {1\over 2}
       \left( E_{S}(y,\xi,Q^2) - E_S(-y,\xi,Q^2)\right) \nonumber \\
       && + {\alpha_s(Q^2) \over 2\pi}
           \left[\int^1_x  {dy\over y} P'_{GG}({x\over y}, {\xi\over y})
            - \int^x_{-1}{dy\over y} P'_{GG}({x\over y}, -{\xi\over y})\right] 
        E_{G}(y,\xi,Q^2) \ , 
\end{eqnarray}
where the evolution kernels are,
\begin{eqnarray}
     P'_{SS}(x, \xi) &=& P'_{NS}(x, \xi)\ ,  \nonumber \\
      P'_{SG}(x, \xi) &=& T_F {(x+\xi/2)(1-2x+\xi/2) \over
\xi(1+\xi/2)(1-\xi^2/4)} \ , \nonumber \\
      P'_{GS}(x, \xi) &=& C_F {(x+\xi/2)(2-x+\xi/2) \over x\xi(1+\xi/2)}
\ ,    \nonumber \\
      P'_{GG}(x, \xi) &=& -C_A {(x^2-\xi^2/4) \over x\xi(1-\xi^2/4)}
               \left[1-{\xi\over 1-x+i\epsilon} - {2(1+x^2) \over
  (1+\xi/2)(x-\xi/2) }\right]  \ . 
\end{eqnarray}
For $x<-\xi/2$, the evolution equations are the same as 
Eq. (\ref{ap1}), except $\int^1_x \rightarrow -\int^x_{-1}$. 

\subsection{Evolution of Parton-Helicity Dependent Distributions} 

The helicity-dependent distributions are defined as follows:
The singlet and non-singlet quark distributions are
\begin{equation}
     \tilde E_{S,NS}(x,\xi,\Delta^2,Q^2)
     ={1\over 2} \int {d\lambda \over 2\pi}
          e^{i\lambda x}\langle P'|\bar \psi(-{\lambda \over 2} n)
        \not\! n\gamma_5 \psi({\lambda \over 2} n) |P\rangle \ . 
\end{equation}
The gluon distribution is, 
\begin{equation}
     \tilde E_G(x, \xi,\Delta^2, Q^2)
    = -{i \over 2x} \int { d\lambda\over 2\pi}e^{i\lambda x}
  \langle P'|F^{\mu\alpha}(-{\lambda\over2}n)
       \tilde F^{\nu}_{~\alpha}({\lambda\over2}n)|P\rangle n_\mu n_\nu \ , 
\end{equation}
where $\tilde F^{\alpha\beta} = {1\over 2}\epsilon^{\alpha\beta\gamma\delta}
F_{\gamma\delta}$. It is easy to see that $\tilde E_G(-x)
=\tilde E_G(x)$. 

The evolution of the non-singlet helicity-dependent quark density
is exactly the same as that of the non-singlet 
helicity-independent quark density. 

For the singlet evolution, we consider mixing
between the singlet quark and gluon distributions. For $x>\xi/2$, the 
evolution equations are,
\begin{eqnarray}
        { D_Q\tilde E_{S}(x,\xi,Q^2)\over D\ln Q^2} &= &{\alpha_s(Q^2) \over 2\pi}
          \int^1_x  {dy\over y} \left[ \Delta P_{SS}({x\over y}, {\xi\over y}) \tilde
      E_S(y,\xi,Q^2)
      + 2 n_F  \Delta P_{SG}({x\over y}, {\xi\over y}) \tilde 
    E_G(y,\xi,Q^2)\right] \ ,  \nonumber \\
        { D_G\tilde E_{G}(x,\xi,Q^2)\over D\ln Q^2} 
       &= &{\alpha_s(Q^2) \over 2\pi}
       \int^1_x  {dy\over y} \left[ 
       \Delta P_{GS}({x\over y}, {\xi\over y}) 
       \times{1\over 2} \left(\tilde E_S(y,\xi,Q^2)
      + \tilde E_S(-y,\xi,Q^2)\right) \right.
       \nonumber \\
    && ~~~~~~~~~~~~~ \left. + ~ \Delta P_{GG}({x\over y}, {\xi\over y}) \tilde
         E_G(y,\xi,Q^2)\right] \  , 
\label{sp}
\end{eqnarray}
where the splitting functions are,
\begin{eqnarray}
      \Delta  P_{SS}(x, \xi) &=& P_{NS}(x, \xi)\ ,  \nonumber \\
     \Delta P_{SG}(x, \xi) &=& T_F {x^2 - (1-x)^2 -\xi^2/4 
      \over (1-\xi^2/4)^2}\ ,  \nonumber \\
     \Delta P_{GS}(x, \xi) &=& C_F {1-(1-x)^2 - \xi^2/4 \over
x(1-\xi^2/4)}\ ,     \nonumber \\
      \Delta P_{GG}(x, \xi) &=& C_A {(x^2-\xi^2/4) \over x(1-\xi^2/4)^2}
               \left[1+{4x(1-x) \over x^2-\xi^2/4} + {1+x-\xi^2/2\over
1-x+i\epsilon}\right] \ . 
\end{eqnarray}
Again, when $\xi=0$, the splitting functions are the usual 
spin-dependent Altarelli-Parisi kernel.
For $-\xi/2<x<\xi/2$, the evolution equations are,
\begin{eqnarray}
        { D_Q\tilde E_{S}(x,\xi,Q^2)\over D\ln Q^2} 
       &=& {\alpha_s(Q^2) \over 2\pi}
       \left[\int^1_x  {dy\over y} \Delta P'_{SS}({x\over y}, {\xi\over
y})  =   - \int^x_{-1}{dy\over y} \Delta P'_{SS}({x\over y}, -{\xi\over y})\right] 
        \tilde E_{S}(y,\xi,Q^2) \nonumber \\
       && + {\alpha_s(Q^2) \over 2\pi} 2n_f
           \left[\int^1_x  {dy\over y} \Delta P'_{SG}({x\over y}, {\xi\over y})
            - \int^x_{-1}{dy\over y} \Delta P'_{SG}({x\over y}, -{\xi\over y})\right] 
        \tilde E_{G}(y,\xi,Q^2) \ , \nonumber \\
        { D_G\tilde E_{G}(x,\xi,Q^2)\over D\ln Q^2} &=& {\alpha_s(Q^2) \over 2\pi}
           \left[\int^1_x  {dy\over y} \Delta P'_{GS}({x\over y}, {\xi\over y})
            - \int^x_{-1}{dy\over y} \Delta P'_{GS}({x\over y}, -{\xi\over y})\right] 
    \nonumber \\
     &&      ~~~~~~~~~~~~~~~ \times {1\over 2} \left( \tilde E_{S}(y,\xi,Q^2)
   +\tilde E_{S}(-y,\xi,Q^2)\right) \nonumber \\
       && + {\alpha_s(Q^2) \over 2\pi}
           \left[\int^1_x  {dy\over y} \Delta P'_{GG}({x\over y}, {\xi\over y})
            - \int^x_{-1}{dy\over y} \Delta P'_{GG}({x\over y}, -{\xi\over y})\right] 
        \tilde E_{G}(y,\xi,Q^2) \ , 
\end{eqnarray}
where the splitting functions are,
\begin{eqnarray}
     \Delta   P'_{SS}(x, \xi) &=& P'_{NS}(x, \xi) \ , \nonumber \\
     \Delta P'_{SG}(x, \xi) &=& T_F {(x+\xi/2)(-1+\xi/2) \over
\xi(1+\xi/2)(1-\xi^2/4)}\ ,  \nonumber \\
     \Delta P'_{GS}(x, \xi) &=& C_F {(x+\xi/2)^2\over x\xi(1+\xi/2)}   \
,  \nonumber \\
     \Delta P'_{GG}(x, \xi) &=& -C_A 
    {(x^2-\xi^2/4) \over x\xi(1-\xi^2/4)}
               \left[1-{\xi \over 1-x+i\epsilon} - {4x \over
       (1+\xi/2)(x-\xi/2)}\right] \ . 
\end{eqnarray}
For $x<-\xi/2$, the evolution equations are the same as Eq. (\ref{sp}), 
except $\int^1_x \rightarrow -\int^x_{-1}$. 

Of course, in a reasonable range of $Q^2$, the OFPDs
do not evolve dramatically. Thus, a second check on the
DVCS dynamical mechanism is to find small $Q^2$ dependences of
relevant scaling functions. 

\section{Cross Sections and Estimates}

In this section, we calculate the cross section 
for electroproduction of a real photon off a nucleon. As shown in Fig. 4, 
we use $k=(\omega, \vec{k})$ and  $k'=(\omega', \vec{k'})$ 
to denote the four-momenta of 
the intial and final electron, $P = (M,0)$ and $P'=(E',\vec{P'})$ 
the intial and final momenta of the nucleon, $q'=(\nu', \vec{q'})$ 
the momentum of the final photon. The 
differential cross section in the laboratory frame is,
\begin{equation}
     d\sigma = {1\over 4\omega M} |{\cal T}|^2
        (2\pi)^4\delta^4(k+P-k'-P'-q')
        {d^3\vec{k'}\over 2\omega'(2\pi)^3}
        {d^3\vec{P'}\over  2E'(2\pi)^3}
        {d^3\vec{q'}\over  2\nu'(2\pi)^3} \ , 
\end{equation}
where $M$ is the nucleon mass and ${\cal T}$ is the T-matrix 
of the scattering. Integrating over the photon momentum, we find,  
\begin{equation}
     d\sigma = {1\over 4\omega M} |{\cal T}|^2
        { \omega'd\omega' d\Omega_{e'} \over 2(2\pi)^3}
        2\pi \delta((k+P-k'-P')^2) {d^3\vec{P'} \over 2E' 
       (2\pi)^3}  \ , 
\end{equation}
where $\delta$-function reflecting the
photon  on-shell condition, which constrains  
the direction 
and magnitude of the momentum of the recoiling nucleon, $\vec{P}'$,
\begin{equation}
    s + M^2 - 2\left((\nu + M)E'-\vec{q}\cdot \vec{P'}\right) = 0 \ , 
\end{equation}
where $s=(q+P)^2$, $q^\mu=(\nu,\vec{q}) = k^\mu-k'^\mu$. 
Thus the phase space of the recoiling proton is specified
by the solid angle $d\Omega_{P'}$. Note, however, that
for large $s$ and $\nu$, there are two solutions of $|\vec{P'}|$
corresponding to one orientation of $\vec{P}'$. Physically one of them
represents the recoil proton at the backward angle in the center-of-mass
frame. Integrating over the magnitude of
$\vec{P}'$, we get, 
\begin{equation}
    d\sigma = {1\over 32(2\pi)^5 \omega M}
           \omega' d\omega' d\Omega_{e'} d\Omega_{P'}
            {  P'^2\over |P'(\nu+M) - qE'\cos\phi|} |{\cal T}|^2 \ , 
\end{equation}
where $\phi$ is the angle between $\vec{q}$ and $\vec{P'}$, 
and the sum over two possible solutions of $|\vec{P'}|$ is 
implicit.
We choose the $z$-axis to be the direction of the incident electron
and the $x$-axis in the plane formed by the initial 
and final electron momenta. In this coordinate system, 
the final electron has the polar angle $\theta$.  
The polar and azimuthal angles of $\vec{P'}$ are denoted
by $\theta_{P'}$ and $\phi_{P'}$, respectively.
Note that $|{\cal T}|^2$ has the dimension of a cross section.

Alternatively, we can integrate out the momentum of the recoil
proton,
\begin{equation}
    d\sigma = {1\over 32(2\pi)^5 \omega M}
           \omega' d\omega' d\Omega_{e'} d\Omega_{q'}
            {  \nu'\over |\nu+M- q\cos\phi'|} |{\cal T}|^2 \ , 
\end{equation}
where $\phi'$ is the angle between $\vec{q}$ and $\vec{q'}$. 
The polar and azimuthal angles of $\vec{q'}$ are denoted
by $\theta_{q'}$ and $\phi_{q'}$, respectively. The constraint
between the energy and direction of the outgoing photon is,
\begin{equation}
    s - M^2 - 2\nu'(\nu + M-q\cos\phi') = 0 \ . 
\end{equation}

\subsection{T-matrix}

We calculate the Feynman diagrams shown in Fig. 4.
We assume for the moment
that the scattering lepton is negatively charged ($e>0$). The T-matrix 
for the Compton scattering part is, 
\begin{equation}
     {\cal T}_1  = - e^3\bar u(k')\gamma^\mu u(k) 
    {1\over q^2} T_{\mu\nu} \epsilon^{\nu *}\ , 
\end{equation}
where $\bar u, u$ are the spinors of the lepton and $\epsilon$ 
is the polarization of the emitting photon. $T_{\mu\nu}$ is the standard
Compton amplitude, 
\begin{equation}
       T_{\mu\nu} = i \int e^{-iq\cdot z} 
 \langle P'|TJ_\mu(z)J_\nu(0)|P\rangle d^4z\ . 
\end{equation}
In the deeply-virtual region, $T_{\mu\nu}$ is expressed
in terms of off-forward parton distributions in Eq. (4). 

The T-matrix for the Bethe-Heitler (BH) process is,
\begin{equation}
       {\cal T}_2 = -e^3 \bar u(k')\left[\not\! \epsilon^*{1\over 
      \not\! k -\not\!\Delta-m_e +i\epsilon}\gamma^\mu + 
     \gamma^\mu {1\over \not\! k' + \not\! \Delta - m_e +i\epsilon}
     \not\!\epsilon^*\right]u(k) {1\over \Delta^2} \langle P'
    |J_\mu(0)|P\rangle \ , 
\end{equation}
where $m_e$ is the mass of electron and will be ignored for the following
discussion. The elastic nucleon matrix element is, 
\begin{equation}
   \langle P'|J_\mu(0)|P\rangle = \bar U(P')\left[\gamma_\mu F_1(\Delta^2) + 
     F_2(\Delta^2) {i\sigma_{\mu\nu} \Delta^\nu \over 2M} \right] U(P) \ , 
\end{equation}
where $\bar U, U$ are the nucleon spinors and
$F_1$ and  $F_2$ are the usual Dirac and Pauli form factors of the nucleon. 

The total T-matrix is the sum of the two above, 
\begin{equation}
      { \cal T }= {\cal T}_1 + {\cal T}_2 \ . 
\end{equation}

\subsection{Unpolarized Scattering}

First, we consider the scattering process without polarizations. 
The square of the T-matrix has three terms. First,
the pure Compton process gives,
\begin{equation}
       |{\cal T}_1|^2 =  - {e^6\over q^4} \ell^{\mu\nu}_{\rm VC}
   W_{\rm VC\mu\nu} \ , 
\end{equation}
where the lepton tensor is, 
\begin{equation}
         \ell^{(\mu\nu)}_{\rm VC} = 2(k^\mu k'^\nu 
       + k^\nu k'^\mu - g^{\mu\nu}k\cdot k')\ . 
\end{equation}
The hadron tensor is calculated from Eq. (4),  
\begin{eqnarray}
             W^{(\mu\nu)}_{\rm VC} &=& {1\over 4} (g^{\mu\nu} - p^\mu n^\nu - p^\nu n^\mu) \Big[
                 \int^1_{-1} dx \alpha(x)\int dx' \alpha^*(x') \nonumber \\
         && [-\xi^2(H(x,\xi,\Delta^2) + E(x, \xi, \Delta^2))
         (H(x',\xi,\Delta^2)+E(x', \xi, \Delta^2)) \nonumber \\ &&
        + 4H(x, \xi, \Delta^2)H(x', \xi, \Delta^2) - {\Delta^2 \over M^2}
          E(x, \xi, \Delta^2) E(x', \xi, \Delta^2)]  \nonumber \\ &&
      + \int^1_{-1} dx \beta(x)\int dx' \beta^*(x') [-\xi^2(\tilde H(x,\xi, 
      \Delta^2) + \tilde E(x, \xi, \Delta^2))
      (\tilde H(x', \xi,\Delta^2) + \tilde E(x', \xi, \Delta^2))  
         \nonumber  \\ && 
       + 4\tilde H(x, \xi, \Delta^2)\tilde H(x', \xi, \Delta^2)
       +\xi^2 \tilde E(x, \xi,\Delta^2) \tilde E(x',
       \xi,\Delta^2)(1-{\Delta^2\over 4M^2}) ]\Big] \ ,
\label{42} 
\end{eqnarray}
where $\alpha(x)$ and $\beta(x)$ are, 
\begin{eqnarray}
    \alpha(x) &=& {1\over x - \xi/2 + i\epsilon} 
               +  {1\over x + \xi/2 - i\epsilon} \ ,  \nonumber \\
    \beta(x) &=& {1\over x - \xi/2 + i\epsilon} 
               -  {1\over x + \xi/2 - i\epsilon} \ .  
\end{eqnarray}
Thus, for $H(x)$ and $E(x)$ less singular than $x^{-2}$ and
$\tilde H(x)$ and $\tilde E(x)$ less singular than $x^{-1}$, 
the integrals are convergent.
To simplify the expression, we have left out as usual 
the sum over quark flavors weighted by the electric charge squared.
To find the product of the two tensors, one  
shall express the null-vectors $n$ and 
$p$ in terms of physical vectors $\bar P$ and $q$ by inverting
Eq. (3).
The result is that $|{\cal T}_1|^2$ goes like
$1/Q^2$ at large $Q^2$, like the inclusive 
deep-inelastic process. 

The pure Bethe-Heitler process gives,
\begin{equation}
      |{\cal T}_2|^2 = {e^6 \over \Delta^4} 
  \ell^{(\mu\nu)}_{BH} W_{\rm BH(\mu\nu)}\ , 
\end{equation}
where the lepton tensor is
\begin{eqnarray}
       \ell^{(\mu\nu)}_{\rm BH} & = &{8\over 
      (k'+\Delta)^4} [k \cdot (k'+\Delta)]
          (2k'^\mu k'^\nu +\Delta^\mu k'^\nu + \Delta^\nu k'^\mu -
      g^{\mu\nu}k'\cdot \Delta) \nonumber \\  && +
     {8\over (k-\Delta)^4} [k' \cdot (k-\Delta)]
          (2k^\mu k^\nu -\Delta^\mu k^\nu - \Delta^\nu k^\mu +
      g^{\mu\nu}k\cdot \Delta) \nonumber \\ &&
      + {8\over (k-\Delta)^2(k'+\Delta)^2}\Big[[2(k'+\Delta)\cdot
      (k-\Delta) + \Delta^2](k'^\mu k^\nu
       + k'^\nu k^\mu) + 2(k'^\mu k'^\nu
     \Delta\cdot k \nonumber \\ && -k^\mu k^\nu \Delta\cdot k')
       + [(k^\mu -k'^\mu)\Delta^\nu - (k'^\nu - k^\nu)\Delta^\mu)
  + (k'-k)\cdot \Delta g^{\mu\nu}]k\cdot k'\Big] \ .
\end{eqnarray}
As $\Delta^\mu\rightarrow 0$, it diverges quadratically as expected
from low-energy theorems.  
The hadron tensor is,
\begin{eqnarray}
        W^{(\mu\nu)}_{\rm BH} & =&  (F_1(\Delta^2) + F_2(\Delta^2))^2 
          (\Delta^2 g^{\mu\nu} 
           - \Delta^\mu\Delta^\nu)  \nonumber \\ &&
        + (P+{\Delta\over 2})^\mu (P+{\Delta\over 2})^\nu 4 
       (F_1(\Delta^2)^2 
       - {\Delta^2\over 4M^2}F_2(\Delta^2)^2) \ ,
\end{eqnarray}
which is well-known from elastic scattering.
Although the product of the two tensors is complicated
in general, it simplifies to
\begin{equation}
         |{\cal T}_2|^2 = {8e^6M^2Q^2\over \Delta^4}
         \left( {\omega\over k\cdot \Delta} - {\omega' \over k'\cdot 
        \Delta}\right)^2 \ . 
\end{equation}
as $\Delta^\mu \rightarrow 0$.
Thus, for small $\Delta^\mu$, the cross section of
the Bethe-Heitler process 
dominates that of DVCS. 
To have a clear DVCS signal, one must have  a
$\Delta^2$ reasonably large (at least on the order of 
the nucleon mass) and a $Q^2$ not too large. 

To appreciate the relative and absolute  
sizes of the DVCS and BH cross sections, 
we shown in Fig. 5 some calculations at the electron beam
energies $\omega$ = 6 GeV (CEBAF after upgrading) 
and $\omega$ = 30 GeV (DESY). For the Bethe-Heitler
process, the cross section can be evaluated accurately 
using the experimentally-measured nucleon form factors.
For the DVCS cross section, we need a model for the OFPDs. Since 
at the moment we are interested in only a rough estimate, we assume 
$E=\tilde E = 0$ and 
\begin{equation}
     H(x, \xi, \Delta^2) = \tilde H(x, \xi,\Delta^2) = 
       q(x) e^{\Delta^2/(2 {\rm ~GeV^2})}\ . 
\end{equation}
In Fig. 5a, we have shown the DVCS (solid) and
BH (dashed) cross sections at incident electron
energy $\omega=6$ GeV and scattering angle 
$\theta=18^\circ$, with a virtual photon $\nu= 3$ GeV and 
$Q^2 = 1.76$ GeV$^2$. 
The recoil proton is detected in the electron scattering plane and 
at the same side of the scattered electron. 
In Fig. 5b, we have shown similar cross sections 
at $\omega = 30$ GeV and scattering
angle $\theta = 3.5^\circ$, with the photon energy
$\nu=7 $ GeV, $Q^2= 2.6 $ GeV$^2$. 
Since the BH cross section is calculable, experimental
cross sections tell us about the DVCS cross section plus
the interference cross section which we now turn to. 

The interference contribution has the following structure, 
\begin{equation}
         {\cal T}_1^*{\cal T}_2 + {\cal T}_1{\cal T}_2^*
     = -2{e^6\over \Delta^2 q^2}\left(\ell^{(\mu\nu)
         \alpha}~{\rm Re} H_{(\mu\nu)\alpha} + 
      \ell^{[\mu\nu]\alpha}~{\rm Re} H_{[\mu\nu]\alpha}\right)\ , 
\end{equation}
where the lepton tensor has been divided into 
symmetric and antisymmetric parts according to the 
indices $\mu$ and $\nu$. The symmetric lepton tensor is
\begin{eqnarray}
       \ell^{(\mu\nu)\alpha} = {2\over (k-\Delta)^2}\Big[&&
        (k'^\mu k^\nu + k'^\nu k^\mu)(k-\Delta)^\alpha
       + [k'^\mu (k-\Delta)^\nu + k'^\nu (k-\Delta)^\mu]k^\alpha 
       \nonumber \\ &&
        + (k'^\mu g^{\nu\alpha} + k'^\nu g^{\mu\alpha})(k\cdot \Delta)
        \nonumber \\ &&
         -g^{\mu\nu}[k\cdot k'(k-\Delta)^\alpha + k'\cdot(k-\Delta)k^\alpha
        + k'^\alpha(k\cdot \Delta) ]\Big] 
        \nonumber  \\
        + {2\over (k'+\Delta)^2}\Big[&&
        (k'^\mu k^\nu + k'^\nu k^\mu)(k'+\Delta)^\alpha
       + [k^\mu (k'+\Delta)^\nu + k^\nu (k'+\Delta)^\mu]k'^\alpha \nonumber \\ &&
        - (k^\mu g^{\nu\alpha} + k^\nu g^{\mu\alpha})(k'\cdot \Delta)
        \nonumber \\ &&
         -g^{\mu\nu}[k\cdot k' (k'+\Delta)^\alpha + k\cdot(k'+\Delta)k'^\alpha
        - k^\alpha(k'\cdot \Delta)] \Big] \ , 
\end{eqnarray}
and the antisymmetric lepton tensor is
\begin{eqnarray}
       \ell^{[\mu\nu]\alpha} &=& {2\over (k-\Delta)^2} \Big[
         -(k\cdot k')(\Delta^\mu g^{\nu\alpha}- \Delta^\nu g^{\mu\alpha})
    + (k'\cdot \Delta)(k^\mu g^{\nu\alpha}-k^\nu g^{\mu\alpha}) \nonumber \\ &&
      - k'^\alpha(k^\mu\Delta^\nu-k^\nu\Delta^\mu) \Big]
       + {2\over (k'+\Delta)^2}\Big[
         (k\cdot k')(\Delta^\mu g^{\nu\alpha}- \Delta^\nu g^{\mu\alpha})\nonumber \\ &&
   - (k\cdot \Delta)(k'^\mu g^{\nu\alpha}-k'^\nu g^{\mu\alpha})
     + k^\alpha(k'^\mu\Delta^\nu-k'^\nu\Delta^\mu)\Big] \ . 
\end{eqnarray}
As $\Delta^\mu \rightarrow 0$, the symmetric part dominates
over the antisymmetric part. 

The hadron tensor is also separated into two contributions.
The symmetric part is,  
\begin{eqnarray}
        H^{(\mu\nu)\alpha} &=& (g^{\mu\nu} - p^\mu n^\nu- p^\nu n^\mu)
\int^1_{-1} dx 
     \alpha^*(x) \nonumber \\ &&
    \Big[ (F_1(\Delta^2)+F_2(\Delta^2))(E(x, \xi,\Delta^2)
       +H(x, \xi, \Delta^2))~[{\xi\over 2}\Delta^\alpha 
     + {\Delta^2\over 2}n^\alpha]
         \nonumber \\ &&
           +   2\bar P^\alpha[H(x,\xi,\Delta^2)F_1(\Delta^2)
           - {\Delta^2\over 4M^2}E(x, \xi,\Delta^2)F_2(\Delta^2)]\Big] \
, 
\end{eqnarray}
and the antisymmetric part is 
\begin{equation}
       H^{[\mu\nu]\alpha} = -\epsilon^{\mu\nu\rho\sigma}n_\rho p_\sigma
       \epsilon^{\beta\gamma\delta\alpha}n_\beta
      p_\gamma \Delta_\delta \int^1_{-1} dx \beta^*(x) 
     (F_1(\Delta^2) + F_2(\Delta^2))\tilde H(x, \xi, \Delta^2) \ . 
\end{equation}
The final expression for products of the tensors above
is quite lengthy and is omitted.
It is generally believed that 
the interference contribution has a size between the BH and DVCS
cross sections. Thus, if the BH cross section is 
much larger than that of DVCS, 
the DVCS amplitude might be accessible through the interference
term which, for instance, can be extracted by comparing
electron and positron scatterings, or by a direct subtraction.

\subsection{Double Spin Process}

In this subsection, we consider scattering with 
both lepton beam and nucleon target polarized.
The structure of the cross section is similar to 
that of the unpolarized case. Indeed, the spin-dependent part of the
${T}$-matrix squared has three contributions. 
The pure virtual Compton process gives,
\begin{equation}
       |{\cal T}_1|^2 =  - {e^6\over q^4} \ell^{[\mu\nu]}_{\rm VC}
     W_{\rm VC[\mu\nu]} \ . 
\end{equation}
The spin-dependent lepton tensor is antisymmetric, 
\begin{equation}
         \ell^{[\mu\nu]}_{\rm VC} = -2\lambda i
         \epsilon^{\mu\nu\alpha\beta}k_\alpha k'_\beta \ , 
\end{equation}
where $\lambda = \pm 1$ represent the positive or negative 
helicity of the scattering lepton. 
The spin-dependent hadron tensor is also antisymmetric,
\begin{eqnarray}
    W^{[\mu\nu]}_{\rm VC} &=& {i\over 2}\epsilon^{\mu\nu\alpha\beta}p_\alpha
            n_\beta  \int dx \int dx' {\rm Re}[\alpha(x) \beta^*(x')] 
     \Big[4 H(x, \xi,\Delta^2) \tilde H(x', \xi, \Delta^2)
   (S\cdot n)(1-{\xi\over 2})  \nonumber \\ &&
         -2E(x, \xi,\Delta^2)\tilde H(x', \xi,\Delta^2)
       [(S\cdot n)(\xi-{\Delta^2 \over 2M^2})-{(S\cdot \Delta)\over M^2}
       (1+{\xi \over 2})]
          \nonumber \\ &&
      - (H(x,\xi,\Delta^2)+E(x,\xi,\Delta^2))\tilde E(x',\xi,\Delta^2) 
      {\xi\over M^2}[(S\cdot \Delta)(1+{\xi\over 2})+(S\cdot n)
     {\Delta^2\over 2}]  \nonumber \\ &&
 +E(x, \xi,\Delta^2)\tilde E(x',\xi,\Delta^2) 
       {\xi\over M^2}(S\cdot \Delta)\Big] \ , 
\end{eqnarray}
where $S^\mu$ is the polarization of the nucleon with normalization
$S^2=-M^2$.
Using $\epsilon^{\mu\nu\alpha\beta}\epsilon_{\mu\nu}^{~~~\gamma\delta}
= -2(g^{\alpha\gamma} g^{\beta\delta}-
g^{\alpha\delta}g^{\beta\gamma})$,
one can straightforwardly 
work out the product of the two tensors. 

The pure Bethe-Heitler process gives,
\begin{equation}
      |{\cal T}_2|^2 = {e^6 \over \Delta^4} \ell^{[\mu\nu]}_{BH} W_{\rm BH [\mu\nu]}\ . 
\end{equation}
The antisymmetric lepton tensor is
\begin{eqnarray}
       \ell^{[\mu\nu]}_{\rm BH} & = & 
       -8\lambda i\epsilon^{\mu\nu\alpha\beta}\Delta_\alpha
         \nonumber \\ && 
         \times \left[{k'_\beta (k'+\Delta)\cdot k\over (k'+\Delta)^4}
       + {k_\beta (k-\Delta)\cdot k'\over (k-\Delta)^4} 
      - {(k+k')_\beta (k\cdot k')
         \over (k-\Delta)^2(k'+\Delta)^2}\right] \ ,
\end{eqnarray}
and the antisymmetric hadron tensor is,
\begin{eqnarray}
        W^{[\mu\nu]}_{\rm BH} &=& i\epsilon^{\mu\nu\alpha\beta}
           \Delta_\alpha\Big[ 2(F_1(\Delta^2)+F_2(\Delta^2))~
    (F_1(\Delta^2)+{\Delta^2\over 4M^2}F_2(\Delta^2))~S_\beta \nonumber \\ &&
       +  F_2(\Delta^2)(F_1(\Delta^2)+F_2(\Delta^2))
         {(S\cdot\Delta)\over M^2}~P_\beta \Big] \ . 
\end{eqnarray}
As $\Delta^\mu\rightarrow 0$, the lepton tensor is subleading 
relative to the spin-independent counterpart shown in Eq. (47).
Thus the spin asymmetry in the BH process vanishes in such a limit.
 
Finally, we consider the interference contribution, 
\begin{equation}
         {\cal T}_1^*{\cal T}_2 + {\cal T}_1{\cal T}_2^*
     = -2{e^6\over \Delta^2 q^2}
    \left(\Delta\ell^{(\mu\nu)\alpha} ~{\rm Re} [\Delta H_{(\mu\nu)\alpha}] + 
      \Delta\ell^{[\mu\nu]\alpha}~{\rm  Re}[\Delta
H_{[\mu\nu]\alpha}]\right) \ , 
\end{equation}
where the spin-dependent lepton tensor has both symmetry and 
antisymmetric parts. The symmetric part is, 
\begin{eqnarray}
       \Delta\ell^{(\mu\nu)\alpha} = {2\lambda i\over (k-\Delta)^2}\Big(&&
       k'^\mu \epsilon^{\nu\alpha\rho\sigma} 
   +   k'^\nu \epsilon^{\mu\alpha\rho\sigma} 
   - g^{\mu\nu}\epsilon^{\lambda \alpha\rho\sigma}k'_\lambda
   \Big)\Delta_\rho k_\sigma \nonumber  \\
        + {2\lambda i\over (k'+\Delta)^2}\Big(&&
       k^\mu \epsilon^{\nu\alpha\rho\sigma} 
   +   k^\nu \epsilon^{\mu\alpha\rho\sigma} 
   - g^{\mu\nu}\epsilon^{\lambda \alpha\rho\sigma}k_\lambda
   \Big)\Delta_\rho k'_\sigma \ , 
\end{eqnarray}
and the antisymmetric part is
\begin{eqnarray}
      \Delta \ell^{[\mu\nu]\alpha} &=& {2\lambda\over (k-\Delta)^2} 
         i\epsilon^{\mu\nu\rho\sigma}\left((k-\Delta)^\alpha
        k_\rho + k^\alpha (k-\Delta)_\rho + g^\alpha_{~\rho}k\cdot\Delta\right)
 k'_\sigma \nonumber \\ &&
 - {2\lambda\over (k'+\Delta)^2} 
         i\epsilon^{\mu\nu\rho\sigma}\left((k'+\Delta)^\alpha
        k'_\rho + k'^\alpha (k'+\Delta)_\rho - g^\alpha_{~\rho}k' 
  \cdot\Delta\right) k_\sigma \ . 
\end{eqnarray}
As $\Delta^\mu \rightarrow 0$, the antisymmetric part dominates over 
the symmetric part. 

The spin-dependent hadron tensor is separated into two parts
accordingly. The symmetric part is, 
\begin{eqnarray}
       \Delta H^{(\mu\nu)\alpha} &=& -i(g^{\mu\nu} - p^\mu n^\nu- p^\nu n^\mu)
       \int^1_{-1} dx 
     \alpha^*(x) \nonumber \\ &&
    \Big[ (F_1(\Delta^2)+F_2(\Delta^2))(E(x,\xi,\Delta^2)
       +H(x,\xi,\Delta^2))~ \epsilon^{\alpha\lambda\rho\sigma}
      n_\lambda S_\rho\Delta_\sigma \nonumber \\ &&
    + F_2(\Delta^2)(H(x,\xi,\Delta^2)+E(x,\xi,\Delta^2))~{\bar 
       P^\alpha\over M^2} \epsilon^{\lambda\rho\sigma\tau}
 p_\lambda S_\rho n_\sigma \Delta_\tau \nonumber\\ &&
  - (F_1(\Delta^2)+F_2(\Delta^2))E(x,\xi,\Delta^2) 
    {1\over M^2} \epsilon^{\alpha\lambda\rho\sigma}P_\lambda
 S_\rho \Delta_\sigma \Big] \ , 
\end{eqnarray}
and the antisymmetric part is,
\begin{eqnarray}
       \Delta H^{[\mu\nu]\alpha}& =& i\epsilon^{\mu\nu\rho
          \sigma}n_\rho p_\sigma
       \int^1_{-1} dx \beta^*(x) \nonumber \\ &&
        \Big[(F_1(\Delta^2)+F_2(\Delta^2))\tilde H(x,\xi,\Delta^2)
         \left[(S\cdot n)2\bar P^\alpha -\xi S^\alpha
 -(S\cdot \Delta) n^\alpha\right] \nonumber \\ &&
  - 2 \tilde H(x,\xi,\Delta^2) F_2(\Delta^2) 
       \bar P^\alpha\left[(S\cdot n)(1-{\Delta^2\over 4M^2})
   - {(S\cdot\Delta)\over 2M^2}(1+{\xi\over 2})\right] \nonumber \\ &&
   - (F_1(\Delta^2)+F_2(\Delta^2))\tilde E(x, \xi,\Delta^2) 
      {\xi\over 2M^2} [(S\cdot\Delta)\bar P^\alpha
    + {\Delta^2\over 2}S^\alpha]  \nonumber \\ &&
+  {\xi\over 2M^2} F_2(\Delta^2) \tilde E(x,\xi,\Delta^2)
   (S\cdot\Delta)  \bar P^\alpha \Big] \ . 
\end{eqnarray}
The product of tensors leads to a lengthy expression which 
we again omit. 

\subsection{Single-Spin Process}

The virtual compton amplitude is complex because of
the intermediate quark
propagation. This gives rise to the so-called 
single-spin asymmetry, that is, the cross-section asymmetry 
depending on a single polarization.
For instance, suppose the lepton beam is polarized and the hadron 
is unpolarized, the single-spin asymmetry is proportional to
\begin{equation}
   A_L^e \sim    -2{e^6\over \Delta^2 q^2}\left( \Delta\ell^{(\mu\nu)\alpha}~
{\rm Im} H_{(\mu\nu)\alpha} + \Delta\ell^{[\mu\nu]\alpha}~
{\rm Im} H_{[\mu\nu]\alpha}\right) \ .        
\end{equation}
On the other hand, if the nucleon is polarized and the lepton is
unpolarized, the single-spin asymmetry is
proportional to 
\begin{equation}
    A^N_{L,T} \sim  -2{e^6\over \Delta^2 q^2}\left(\ell^{(\mu\nu)\alpha} 
     ~{\rm Im} [\Delta H_{(\mu\nu)\alpha}] + 
    \ell^{[\mu\nu]\alpha} ~{\rm Im}[\Delta H_{[\mu\nu]\alpha}]\right)
\ .
\end{equation}
The size of these spin asymmetries directly reflect the relative
contributions of the DVCS and BH processes. Their measurement
and interpretation are very interesting although they only depend
on the parton distributions at $x=\pm \xi/2$. 

\section{Comments and Conclusions}

In this paper, we studied some basic aspects of 
deeply-virtual Compton scattering. The motivation is
that in the deeply-virtual kinematics, the scattering
mechanism appears to be simple, and hence, one can
learn some structural information from the process. 
The QCD analysis shows that it is the off-forward
parton distributions that are probed. In a particular
experiment, of course, one has to decide whether one
is at the deeply-virtual kinematics. As we discussed
in the paper, one can look at certain observables which 
vanish in the kinematic limit (like $R=\sigma_L/\sigma_T$
in deep-inelastic scattering). Those include 
$\sigma_L$, $\sigma_{TT}$ and $\sigma_{LT}$ discussed in
Ref. \cite{guichon}. One can also check the
slow-$Q^2$ dependence of certain scaling functions. 
As a first guess, one shall be at least 
at the deep-inelastic kinematic region, and the 
recoil nucleon shall be at backward angles.    
It would be interesting to 
demonstrate experimentally that the single-quark
scattering mechanism is at work. 

One important theoretical issue is whether a factorization
theorem exists for the deeply-virtual 
Compton process. The answer is most likely yes
for several reasons. First, the tree result we obtained
can be easily cast into an operator product expansion.
Such an expansion is generally believed to be valid
independent of external states. Second, the radiative
corrections to the tree process are quite similar
to the corrections to $\gamma^*\pi\rightarrow \gamma$, 
a process known to be factorizable at one loop level. 
Nonetheless, one still has to prove explicitly 
the factorization theorem
for DVCS, which will be done in a separate 
publication. A factorization proof for a similar 
process, electroproduction of mesons, has appeared
recently \cite{co}.

DVCS at small $\Delta^2$ is especially interesting,
because the nucleon form-factor suppression is small
and because $\Delta^2 \rightarrow 0$ limit is relevant
to the spin structure of the nucleon. However,
due to QED infrared divergences, the Bethe-Heitler 
process becomes dominant there despite the fact 
that the DVCS cross section
scales like the deep-inelastic scattering cross section. 
Thus one cannot 
go to too small $\Delta^2$. One may get around this
to some extent by isolating the DVCS and BH interference term
through single-spin asymmetry and/or combined
lepton-antilepton scattering. Theoretical study of 
$\Delta^2$ dependence of the OFPDs, in particular,
its relation to the meson-dominance and the exponential 
decay law for exclusive cross sections, is urgently needed.

The OFPDs depend on four independent variables:
the Bjorken-type of variable $\xi$, the Feynman-type 
variable $x$, the momentum transfer $\Delta^2$ 
and the virtual-photon mass $Q^2$. 
It is unlikely that one can learn 
the entire parameter space of the distributions 
in one kind of experiments. As our formulas 
show, different kinds of experiments are sensitive
to different combinations of parton distributions
and hence they are complementary. 
[The reason that the nucleon helicity-flip
distributions contribute to unpolarized scattering
is similar to that the Pauli-form factor $F_2(\Delta^2)$
appears in unpolarized elastic scattering.] 
However, to be able to study sum rules, one must
have accurate data in an extended kinematic region.
To achieve this, one must have dedicated experiments 
at a suitable machine like ELFE for extended running \cite{elfe}. 

The off-forward parton distributions
can also be defined for quark helicity-flip (chiral-odd)
correlations and for higher twists. 
DVCS provides
one process to access to these distributions. There
are other processes one can consider to measure
them. For instance, the diffractive $\rho$ or $J/\psi$ 
production studied by Brodsky et al.\cite{brodsky}
can be used to measure the off-forward gluon distributions. 
Recently, Radyushkin has published a paper aimed 
at this direction\cite{r2}. 
Thus there is now a new territory to explore 
the quark and gluon structure of the nucleon besides 
the traditional inclusive (parton distributions)
and exclusive (form factors) processes.

\acknowledgements
I thank I. Balitksy, D. Beck, V. Braun, S. Brodsky, R. Jaffe, 
R. McKeown, A. 
Nathan and A. Radyushkin for their discussions and interest, 
S. Forte, P. Guichon and A. Radyushkin for their constructive 
comments and criticisms about the manuscript, 
and O. Teryaev for pointing out reference [\cite{geyer}].

\begin{figure}
\caption{Virtual Compton scattering process.}
\label{fig1}
\bigskip
\caption{QCD diagrams for deeply-virtual Compton scattering:
a). the hand-bag diagram, b). \& c). some $1/Q^2$ corrections, 
d). \& e). some radiative corrections.}
\label{fig2}
\bigskip
\caption{Leading-log evolution of the off-forward 
parton distributions.}
\label{fig3}
\bigskip
\caption{Electroproduction of a photon off the nucleon: a). the
virtual Compton scattering, b). \& c). the Bethe-Heitler process.}
\label{fig4}
\bigskip
\caption{Comparison of the DVCS and BH cross sections at a). CEBAF
and b). HERMES kinematics.}
\label{fig5}
\end{figure}

\begin{references}
\frenchspacing

\bibitem{compton}
A. H. Compton, Phys. Rev. 22 (1923) 409.

\bibitem{low}
F. Low, Phys. Rev. 96 (1954) 1428; M. Gell-Mann and M. L. Goldberger, 
Phys. Rev. 96 (1954) 1433.

\bibitem{pol1}
See for instance, V. B. Berestetskii, E. M. Lifshitz and L. P. 
Pitaevskii, {\it Quantum Electrodynamics}, Pergamon Press, 1980.  

\bibitem{pol2}
F. J. Federspiel et al., Phys. Rev. Lett. 67 (1991) 1511;
A. Zieger et al., Phys. Lett. B 278 (1992) 34; E. Hallin et al., 
Phys. Rev. C 48 (1993) 1497.

\bibitem{farrar}
G. R. Farrar, G. Sterman, and H. Zhang, Phys. Rev. Lett. 62 (1989) 2229;
G. R. Farrar and H. Zhang, Phys. Rev. D41 (1990) 3348. 

\bibitem{brodsky0}
S. J. Brodsky and G. R. Farrar, Phys. Rev. Lett. 31 (1973) 1153. 

\bibitem{e143}
J. Ashman et al., Nucl. Phys. B328 (1989) 1; B. Adeva et al., Phys. 
Lett. B302 (1993); P. L. Anthony et al., Phys. Rev. Lett. 71 (1993)
959; K. Abe et. al., Phys. Rev. Lett. 75 (1995) 25. 

\bibitem{ji}
X. Ji, Phys. Rev. Lett. 78 (1997) 610. 

\bibitem{r1}
A. V. Radyushkin, Phys. Lett. B380 (1996) 417. 

\bibitem{geyer}
F. M. Dittes, D. Muller, D. Robsschik, B. Geyer, and J. Horejsi,
Phys. Lett. B 209 (1988) 325; Fortschr. Phys. 42 (1994) 101;
B. Geyer, D. Robaschik, M. Bordag, and J. Horejsi, 
Z. Phys. C 26 (1985) 591; T. Braunshweig, 
B. Geyer, J. Horejsi, and D Robaschik, Z. Phys. C 33 (1987) 275;

\bibitem{ral}
P. Jain and J. P. Ralston, in the proceedings of the workshop
on Future Directions in Particle and Nuclear Physics at Multi-GeV 
Hadron Beam Facilities, BNL, March, 1993.

\bibitem{guichon}
P. Kroll, M. Sch\"urmann, and P. A. M. Guichon,
Nucl. Phys. A598 (1996) 435.

\bibitem{jaffeji}
R. L. Jaffe and X. Ji, Phys. Rev. Lett. 67 (1991) 552.

\bibitem{ap}
G. Altarell and G. Parisi, Nucl. Phys. B126 (1977) 278. 

\bibitem{braun}
I. Balitsky and V. Braun, Nucl. Phys. B311 (1989) 541.

\bibitem{r2}
A. V. Radyushkin, Hep-ph/9605431, CEBAF-TH-96-06, May, 1996.

\bibitem{jaffe0}
R. L. Jaffe, Nucl. Phys. B229 (1983) 205. 

\bibitem{bl}
G. P. Lepage and S. J. Brodsky, Phys. Rev. D22 (1980) 2157;

\bibitem{co}
J. C. Collins, L. Frankfurt, and M. Strikman, hep-ph/9611433, 1996.

\bibitem{elfe}
J. Arvieux and E. de Sanctis, The ELFE project, Italian 
Physical Society Conference Proceeding 44 (1993).  

\bibitem{brodsky}
S. J. Brodsky, L. Frankfurt, J. F. Gunion, A. H. Mueller, M. Strikman, 
Phys. Rev. D50 (1994) 3134.  


\nonfrenchspacing
\end{references}
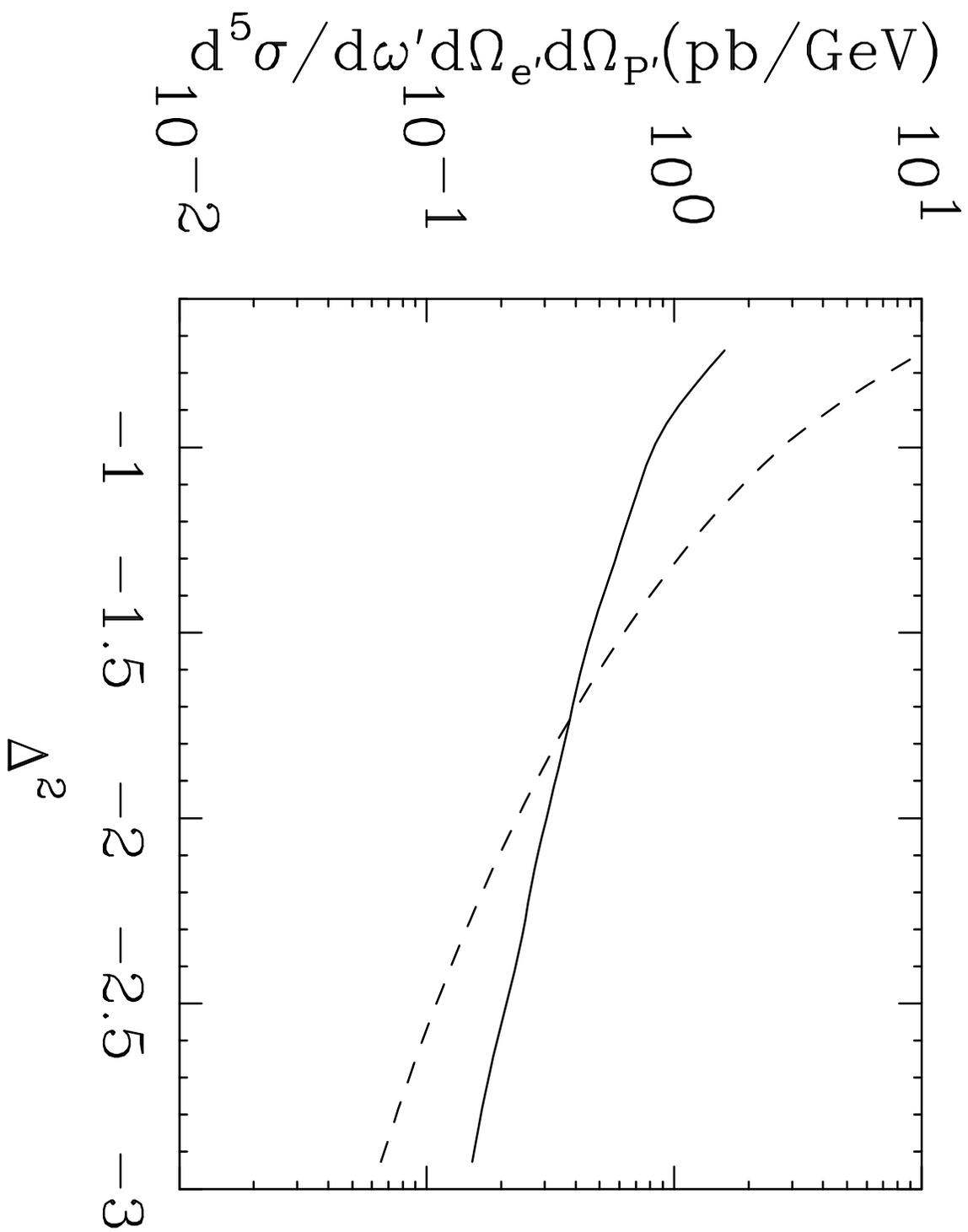
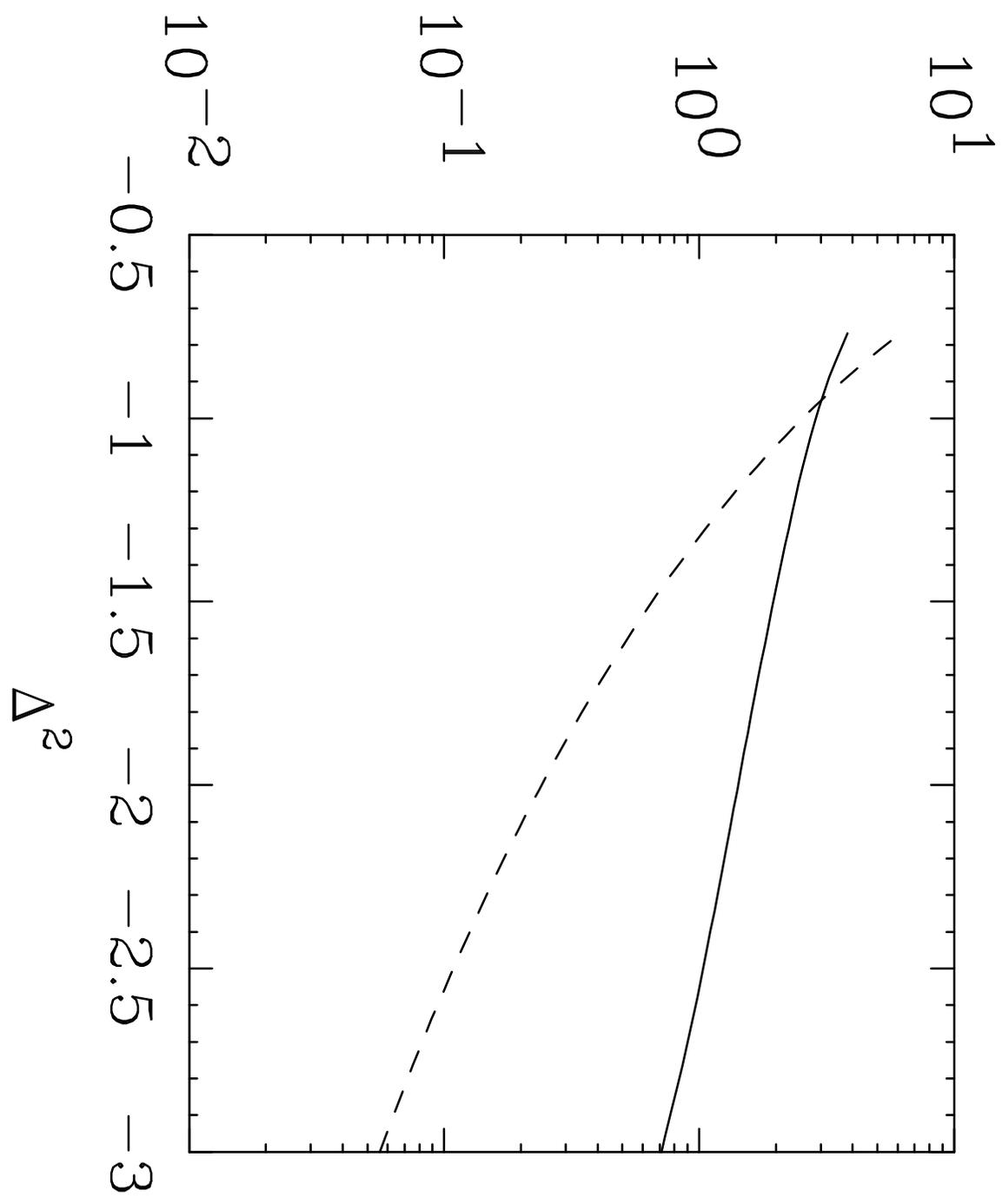
\end{document}